\documentclass[12pt]{iopart}

\newcommand{\cecoin}    {CeCoIn$_5$}
\newcommand{\cerhin}    {CeRhIn$_5$}

\newcommand{\cemin}    {CeMIn$_5$}
\newcommand{\pucoga}    {PuCoGa$_5$}

\newcommand{\pumga}    {PuMGa$_5$}

\newcommand{\tc}     {$T_{\rm c}$}
\newcommand{\slrr}  {$T_1^{-1}$}
\newcommand{\hhyp}  {$H_{\rm hyp}$}

\newcommand{\slrrtext}  {spin lattice relaxation rate}

\usepackage{iopams}
\usepackage{graphicx}
\begin{document}


\title[Hyperfine Couplings in CeMIn$_5$]{Hyperfine Interactions in the Heavy Fermion CeMIn$_5$ Systems}

\author{N. J. Curro}

\address{Condensed Matter and Thermal Physics, Los Alamos
National Laboratory, Los Alamos, NM 87545, USA} \ead{curro@lanl.gov}
\begin{abstract}
The \cemin\ heavy fermion compounds have attracted enormous interest
since their discovery six years ago.  These materials exhibit a rich
spectrum of unusual correlated electron behavior, and may be an
ideal model for the high temperature superconductors.  As many of
these systems are either antiferromagnets, or lie close to an
antiferromagnetic phase boundary, it is crucial to understand the
behavior of the dynamic and static magnetism. Since neutron
scattering is difficult in these materials, often the primary source
of information about the magnetic fluctuations is Nuclear Magnetic
Resonance (NMR). Therefore, it is crucial to have a detailed
understanding of how the nuclear moments interact with conduction
electrons and the local moments present in these systems. Here we
present a detailed analysis of the hyperfine coupling based on
anisotropic hyperfine coupling tensors between nuclear moments and
local moments.  Because the couplings are symmetric with respect to
bond axes rather than crystal lattice directions, the nuclear sites
can experience non-vanishing hyperfine fields even in high symmetry
sites.
\end{abstract}

\pacs{71.27.+a, 75.30.Mb, 76.60.-k}
\submitto{\NJP}
\maketitle

\section{Introduction to NMR in the CeMIn$_5$ Materials}
The \cemin\ materials (also known as the 115's), with M$=$ Co, Rh or
Ir, exhibit a rich variety of phenomena as a result of strong
electron-electron correlation \cite{joereview}. These phenomena
include unconventional d-wave superconductivity, local moment
magnetism, coexistence of magnetism and superconductivity, and a
possible Fulde-Ferrell-Larkin-Ovchinnikov superconducting phase
\cite{romanFFLO,tusonNature}. The superconducting transition
temperature, \tc, of \cecoin\ is the highest of the Ce-based heavy
fermion materials. Recently, an isostructural class of plutonium
based 115's was synthesized, the \pumga\ series, with M$=$ Co or Rh,
that are superconductors with \tc's of 18.5K, and 8.5K, respectively
\cite{pucoga5discovery}. Since \pucoga\ is a d-wave superconductor
like \cecoin, it has been argued that its ten-fold increase in \tc\
is due to an increase in the size of the relevant magnetic
interaction energy, $J$, and thus the \pumga\ materials bridge the
gap between two distinct sets of unconventional superconductors: the
heavy fermion superconductors, with \tc$\sim1-2$K, and the high
temperature superconductors, with \tc$\sim 100$K
\cite{curropucoga5nature}. Consequently, one can reasonably argue
that by investigating the role of magnetism and superconductivity in
the \cemin\ materials, one can gain considerable insight in the the
role of magnetism and superconductivity in the high-\tc\ materials
and perhaps shed new light on the details of the pairing mechanism.
These heavy fermion superconductors allow us to probe the physics
over a much broader temperature range compared to the characteristic
spin fluctuation energy, $J$ (where $J$ typically is on the order of
1500K for the cuprates and 10K for the heavy fermions). Furthermore,
the \cemin\ materials are more easily handled in the laboratory, can
be tuned with pressure rather than doping, and are extremely clean
with small residual resistivities and large scattering lengths.

\begin{figure}
\begin{centering}
\includegraphics[width=0.6\linewidth]{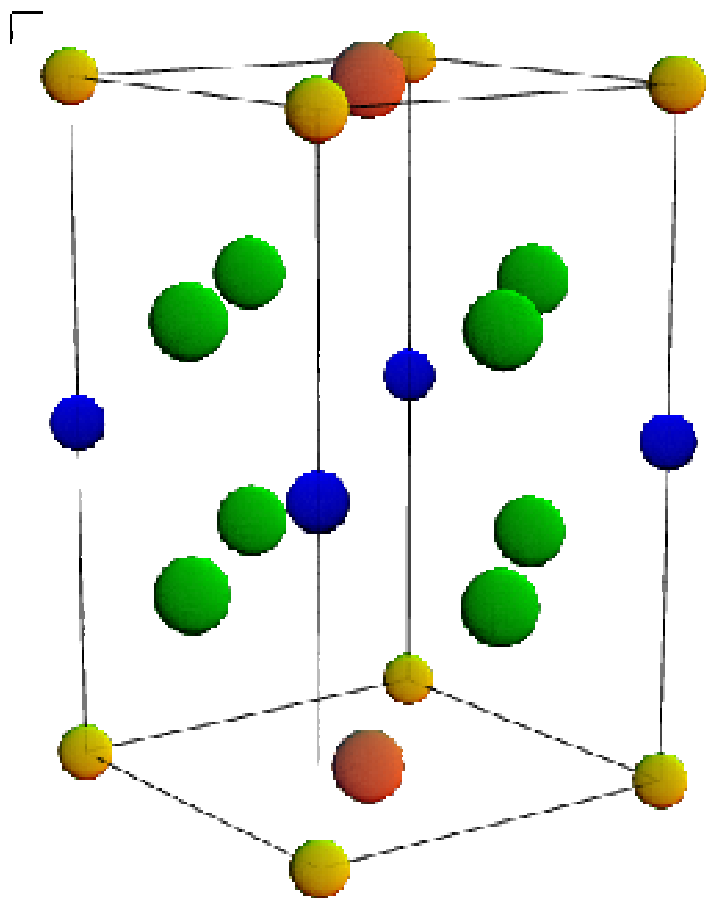}
\caption{\label{fig:115structure} The structure of the \cemin\
materials.  The Ce (Wykyoff position 1a) atoms are yellow, the In(1)
(1c) atoms are orange, the In(2) (4i) are green, and the M atoms are
blue (1b).  Details of the structural parameters can be found in
\cite{structure115}.}
\end{centering}
\end{figure}

The magnetism in the \cemin\ system is clearly relevant at some
level for the existence of the unconventional superconductivity and
the non-Fermi liquid behaviors \cite{sidorov,
bauer,ronning,curroMRS}. The magnetism can be investigated
experimentally by measuring the dynamical magnetic susceptibility
experimentally, either with neutron scattering or NMR.  Neutron
scattering experiments often are limited because the $^{115}$In
strongly absorbs neutrons (cross section $\sim200$ barns), which
severely limits the resolution. On the other hand, NMR can provide
important insight into the behavior of the dynamical susceptibility
in these systems through measurements of the \slrrtext, \slrr, and
the Knight shift, $K$, of the In, Ga or M nuclei. Since the nuclear
spins are coupled to the electron spin moments in the system, they
are sensitive to the dynamics of these electron spins. In fact,
\slrr, is sensitive to the imaginary part of the dynamical electron
spin susceptibility, whereas $K$ probes the static susceptibility.
The details of how the nuclei couples to the electron spins,
however, are important for interpreting the \slrr\ and $K$ data. For
example, if a nucleus is coupled to more than one electron spin,
then the hyperfine field can vanish for particular spin
configurations at certain nuclear sites, depending on the symmetry
of the site.

In this manuscript we attempt to provide a complete analysis of the
hyperfine couplings to the various NMR-active sites in this 115's.
Since their discovery six years ago, several groups have reported
Knight shift and \slrrtext\ data in these materials. However, in
order to make any connection between the NMR data and physically
relevant quantities such as the dynamic or static susceptibility,
one needs to understand the hyperfine interaction. Shortly after the
first NMR experiments in the cuprates, Mila and Rice showed that the
response of the Cu and O can be understood in a relatively simple
single spin component model by elucidating the hyperfine
interactions \cite{milarice}. Their pioneering work led to
significant insight into the cuprates as observed by NMR. The
hyperfine scenario in the 115's, and the heavy fermion compounds in
general, is more complex than that in the cuprates. There is strong
evidence for two components of spin susceptibility, with different
temperature dependences. Since the spin lattice relaxation is
dominated by fluctuating hyperfine fields, these two components can
contribute in unexpected ways to the temperature dependence.
Consequently, one must approach any interpretation of \slrr\ data
with an eye to these anomalous effects, and a complete analysis
requires an intimate understanding of the hyperfine interactions.

\section{Transferred Hyperfine Couplings}

Most of our knowledge of the hyperfine interactions in the 115's
comes from measurements of $K$ and the static hyperfine field,
\hhyp\ in the magnetically ordered state.  In general, a nuclear
spin $\hat{\mathbf{I}}$ is coupled to an electron spin
$\hat{\mathbf{S}}$ via an interaction, $A$:
\begin{equation}
\hat{\mathcal{H}}_{\rm hyp}=\gamma\hbar
A\hat{\mathbf{I}}\cdot\hat{\mathbf{S}},
\end{equation}
where $\gamma$ is the gyromagnetic ratio. In an external field
$H_0$, the electron spins are polarized via  their uniform (i.e.
$\mathbf{q}$=0) susceptibility: $M =
g\mu_B\langle\hat{\mathbf{S}}\rangle = \chi H_0$. This leads to a
shift, $K$, in the NMR resonance frequency:
\begin{equation}
\hat{\mathcal{H}}=\gamma\hbar\hat{\mathbf{I}}\left(1+K\right)\cdot
\mathbf{H}_0,
\end{equation}
where $K= A\chi/g\mu_B$.  If the susceptibility is strongly
temperature dependent, then the Knight shift should be as well.  By
measuring $K(T)$ and $\chi(T)$ independently, and plotting $K$
versus $\chi$ with $T$ as an implicit parameter, one can extract the
value of the hyperfine coupling constant, $A$.

In the \cemin\ materials, $\chi$ is strongly temperature dependent
with Curie-Weiss behavior. There are two sets of spins that
potentially can give rise to this susceptibility: the conduction
electron spins, $S_c=1/2$, and the local moments of the Ce$^{3+}$
4f$^1$ electrons, $S_f$. Note that for the lanthanides, the
spin-orbit interaction is several eV, so for experimentally
realizable temperatures, the important quantum number is the total
spin $J = L + S$, where $L = 3$ and $S=1/2$ for Ce$^{3+}$. The
degeneracy of the ground state 5/2 multiplet is further lifted by
the crystal field interaction, which has tetragonal symmetry in the
115's, typically on the order of 50-100K, so that at low
temperatures the ground state of the Ce$^{3+}$ is a pseudospin
doublet with an effective $g$ value determined by the crystal field.
For simplicity, we denote the 4f spin by $S_f$. The local moments
and the conduction electrons also experience a Kondo interaction and
an exchange interaction between different 4f spins that is driven by
an RKKY or other similar mechanism. The hyperfine interaction
couples the nuclei to these two sets of spins, but the coupling
strength is several orders of magnitude smaller than the other
interactions in the system. Therefore, the nuclei are sensitive
probes of the magnetic behavior of the system, but they do not
significantly perturb the intrinsic phenomena. Empirically, we find
that for sufficiently large temperatures $T>T^*$, where $T^*$ is
material dependent but on the order of 50K, $K$ is linearly
proportional to $\chi$. Since the dominant contribution to the
susceptibility is from the local moments, $S_f$, at these
temperatures, we conclude that the largest hyperfine field arises
from the local moments. In general, however, there can be also an
on-site hyperfine interaction to conduction electrons.  A more
complete description of the hyperfine interactions is given by:
\begin{equation}
\hat{\mathcal{H}}_{\rm hyp}=\gamma\hbar\mathbf{I}\cdot\left(
\mathbb{A}\cdot\hat{\mathbf{S}}_c+ \sum_{i\in
nn}\mathbb{{B}}\cdot\hat{\mathbf{S}}_f\right),
\end{equation}
where $\mathbb{A}$ is an on-site hyperfine tensor interaction to the
conduction electron spin, and $\mathbb{B}$ is a transferred
hyperfine tensor to the Ce 4f spins \cite{twocomponentNMR}.  A
similar scenario is present in the high temperature superconductors
\cite{milarice}. For the \cemin\ materials, the mechanism of the
on-site coupling $\mathbb{A}$ is probably due to a combination of
core polarization, and polarization of unfilled orbitals with s and
p symmetry.  The mechanism of the transferred interaction,
$\mathbb{B}$ is probably due to a combination of orbital overlap
between the 4f wavefunction and ligand nuclei core wavefunctions,
and a polarization of the conduction electron spins via the Kondo
interaction between $\hat{\mathbf{S}}_c$ and $\hat{\mathbf{S}}_f$.
For the remainder of this article, we assume that the hyperfine
parameters are material dependent constants, and will not discuss
their microscopic mechanism further.

\subsection{Coupling to In(1)}

In the antiferromagnetic compound \cerhin\ ($T_N=3.8$K) the magnetic
order is an incommensurate spiral with the moments in the $ab$
plane, and the modulation along the $c$ direction, with
$Q_{\mathbf{AF}} = (0.5,0.5,0.297)$ \cite{curroCeRhIn5,BauCeRhIn5}.
In this system, the In(1) lies in the plane with four nearest
neighbor Ce spins, and by symmetry one might expect the transferred
hyperfine field would vanish in the ordered state.  In fact, the
hyperfine field is finite ($H_{\mathrm{hyp}} \approx 1.8$kOe), and
is about five times larger than one would expect from a direct
dipolar interaction between localized 4f spins and the In ($I=9/2$)
nuclear spin.  The reason the transferred hyperfine does not vanish
is that the eigenvectors of the tensor $\mathbb{B}$ do not coincide
with the unit cell coordinates, but rather point along the bond
directions between the Ce and the In(1).  We can write $\mathbb{B}$
as:
\begin{equation}
\mathbb{B} = \left(
               \begin{array}{ccc}
                 B_{||} & 0 & 0 \\
                 0 & B_{\perp} & 0 \\
                 0 & 0 & B_{c} \\
               \end{array}
             \right)
\end{equation}
in the bond-coordinate system, where $B_{\alpha}$ corresponds to the
transferred hyperfine coupling along the $\alpha$ direction as shown
in Fig. \ref{fig:hyperfinetensor}.  In the unit cell (and
experimental) reference frame, this tensor becomes
\begin{equation}
\mathbb{B}_{1,3} = \left(
               \begin{array}{ccc}
                 B_{0} & B_{a} & 0 \\
                 B_{a} & B_{0} & 0 \\
                 0 & 0 & B_{c} \\
               \end{array}
             \right)
\end{equation}
and
\begin{equation}
\mathbb{B}_{2,4} = \left(
               \begin{array}{ccc}
                 B_{0} & -B_{a} & 0 \\
                 -B_{a} & B_{0} & 0 \\
                 0 & 0 & B_{c} \\
               \end{array}
             \right)
\end{equation}
for the sites $i\in(1,2,3,4)$ in Fig. {\ref{fig:115structure}, where
$B_0 = \left(B_{||} + B_{\perp}\right)/2$ and $B_a = \left(B_{||} -
B_{\perp}\right)/2$.

\begin{figure}
\begin{centering}
\includegraphics[width=0.6\linewidth]{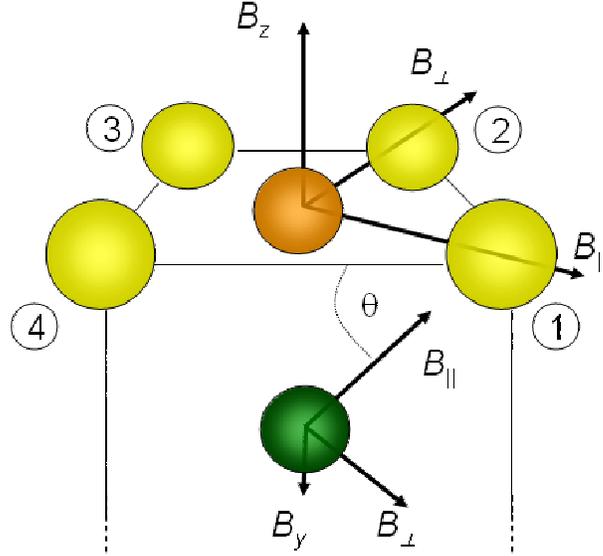}
\caption{\label{fig:hyperfinetensor} The principle axes of the
transferred hyperfine tensors $\mathbb{B}_i$ of the In(1) and the
In(2) sites.}
\end{centering}
\end{figure}

The on-site interaction, $\mathbb{A}$ is probably anisotropic, but
diagonal in the unit cell coordinates:
\begin{equation}
\mathbb{A} = \left(
               \begin{array}{ccc}
                 A_{ab} & 0 & 0 \\
                 0 & A_{ab} & 0 \\
                 0 & 0 & A_c \\
               \end{array}
             \right)
\end{equation}
The Knight shift at the In(1) site, then should be.
\begin{eqnarray}
\label{eqn:KSIn1}
K_{\alpha}(T) &=& \left(A_{\alpha}\langle\hat{S}_c^{\alpha}\rangle+4B_{\alpha}\langle\hat{S}_f^{\alpha}\rangle\right)/H_0\\
\nonumber &\approx& 4B_{0}\chi_{\alpha}(T),
\end{eqnarray}
where the approximation should hold for temperatures $T>T^*$, where
the Curie-Weiss susceptibility of the 4f electrons dominates the
Pauli susceptibility of the conduction electrons.  Below, in section
3, we discuss the Knight shift for $T<T^*$, where the approximation
no longer holds.

In the antiferromagnetic state of \cerhin\ the ordering is given by:
\begin{equation}
\langle\mathbf{S}_f(\mathbf{r})\rangle = S_0\cos\left(\frac{\pi
x}{a}\right)\cos\left(\frac{\pi y}{a}\right)\left(\cos(2\pi
q_0z)\hat{\mathbf{x}} +\sin(2\pi q_0z)\hat{\mathbf{y}}\right),
\end{equation}
where $q_0$ is the wavevector of the spiral along the $c$ direction
\cite{curroCeRhIn5,BauCeRhIn5}. The hyperfine field at the In(1)
site is then given by:
\begin{equation}
\mathbf{H}_{\mathrm{hyp}} = 4B_aS_0\left(\sin(2\pi
q_0z)\hat{\mathbf{x}} +\cos(2\pi q_0z)\hat{\mathbf{y}}\right).
\end{equation}
$H_{\rm hyp}\propto(B_{||} - B_{\perp})$ would vanish if
$\mathbb{B}$ did not have off-diagonal terms in the coordinates of
the unit cell.  Also, since $\mathbf{H}_{\mathrm{hyp}} \perp c$, and
the electric field gradient (EFG) vector for the In(1) is parallel
to $c$, the resonance frequency in the antiferromagnetic state is
independent of $z$. Consequently, even though the magnetism is
incommensurate, the In(1) lines remain sharp \cite{curroCeRhIn5}.
The orientation of the 4f moments and the hyperfine fields are shown
in Fig. \ref{fig:115hyperfine}.

\begin{figure}
\begin{centering}
\includegraphics[width=0.6\linewidth]{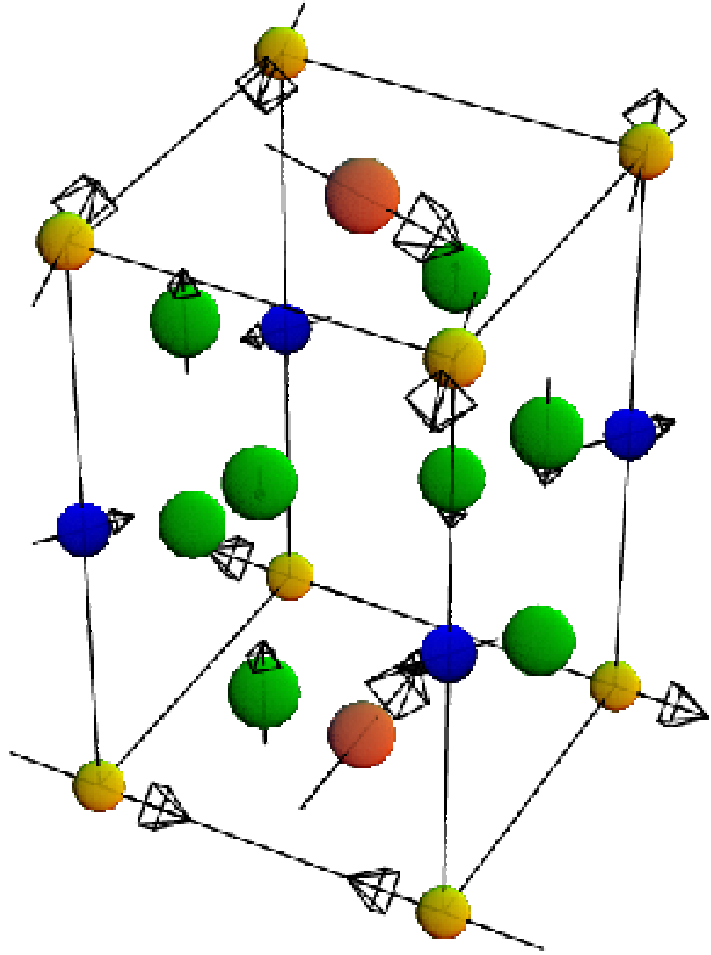}
\caption{\label{fig:115hyperfine} The hyperfine fields in the
\cemin\ materials. The atoms are the same as in Fig.
\ref{fig:115structure}.  The arrows on the Ce atoms denote the
direction of the electronic moments, for the ordered structure in
the \cerhin\ antiferromagnetic state.  The arrows on the other atoms
(In(1), In(2) or M) denote the direction of the hyperfine fields at
the particular site, as discussed in the text.  An animated version
of this figure is available online, showing how the hyperfine fields
change as the direction of the Ce moments changes.}
\end{centering}
\end{figure}

A further consequence of anisotropic transferred hyperfine tensor,
$\mathbb{B}$ is that the In(1) is sensitive to the critical
fluctuations of the order parameter above $T_N$.  We will come back
to this subject later in a discussion of the form factors.

\subsection{Coupling to In(2)}

The In(2) is in a low symmetry site out of the Ce plane.  We expect
that the transferred hyperfine coupling tensor to this site should
also be diagonal along the bond axes (see Fig.
\ref{fig:hyperfinetensor}). In the unit cell coordinate system, we
have then:
\begin{equation}
\mathbb{B}_{1} = \left(
               \begin{array}{ccc}
                 B_{0} + B_a\cos{2\theta} & 0 & B_a\sin{2\theta} \\
                 0 & B_{y} & 0 \\
                 B_a\sin{2\theta} & 0 & B_{0} - B_a\cos{2\theta} \\
               \end{array}
             \right)
\end{equation}
and
\begin{equation}
\mathbb{B}_{2} = \left(
               \begin{array}{ccc}
                 B_{0} + B_a\cos{2\theta} & 0 &- B_a\sin{2\theta} \\
                 0 & B_{y} & 0 \\
                 -B_a\sin{2\theta} & 0 & B_{0} - B_a\cos{2\theta} \\
               \end{array}
             \right)
\end{equation}
where $\theta$ is the angle between the Ce-In(2) direction and the
$a$ axis.  For \cerhin\ $\theta \approx 45^\circ$.  In this case,
the Knight shift is given by:
\begin{eqnarray}
K_a &=& \left(A_{a}\langle\hat{S}_c^{a}\rangle + 2(B_0 +
B_a\cos(2\theta))\langle\hat{S}_f^{a}\rangle\right)/H_0\\
K_b &=& \left(A_{b}\langle\hat{S}_c^{b}\rangle + 2B_y\langle\hat{S}_f^{b}\rangle\right)/H_0\\
K_c &=& \left(A_{c}\langle\hat{S}_c^{c}\rangle + 2(B_0 -
B_a\cos(2\theta))\langle\hat{S}_f^{c}\rangle\right)/H_0,
\end{eqnarray}
where we have ignored the small component of hyperfine field that is
perpendicular to $H_0$. As in the case for In(1), we assume that the
Pauli susceptibility is negligible for $T>T^*$, so
\begin{eqnarray}
\label{eqn:KSIn2}
K_a &\approx&  2(B_0 +B_a\cos(2\theta))\chi_{ab}(T)\\
\nonumber K_b &\approx&  2B_y\chi_{ab}(T)\\
\nonumber K_c &\approx& 2(B_0 -B_a\cos(2\theta))\chi_c(T).
\end{eqnarray}
Even though the spin susceptibility is isotropic in the $ab$ plane,
the In(2) shift depends on the whether the field is in the $a$ or
$b$ directions.  The reason for this is that the symmetry of the
In(2) site is lowered when the field is in the plane, so two of the
In(2) sites have the field applied parallel to the face of the unit
cell, and two of the sites have a field perpendicular to the face of
the unit cell.

 In the Ne\'{e}l state, the hyperfine field at the In(2) site is
parallel to the $c$ axis:
\begin{equation}
\mathbf{H}_{\mathrm{hyp}} = 2B_aS_0\cos(2\pi q_0 z)\hat{\mathbf{z}}.
\end{equation}
Once again, the hyperfine field at the In(2) site is finite only
because the tensor $\mathbb{B}$ is not diagonal in the unit cell
coordinates, and thus the In(2) is also sensitive to critical
fluctuations.  The orientation of the hyperfine field at the In(2)
sites is shown in Fig. \ref{fig:115hyperfine}.  Note that in the
case of the In(2), the resonance frequency is dependent on the $z$
coordinate, therefore the In(2) exhibits a broad "powder
pattern"-type resonance spectrum in the ordered state of \cerhin\
\cite{curroCeRhIn5}.

\subsection{Coupling to M}

For the M site, the bond axes coincide with the unit cell axes, so
the hyperfine tensor is diagonal in the unit cell basis.  We thus
have:
\begin{equation}
\mathbb{B}_{1,2} = \left(
               \begin{array}{ccc}
                 B_{ab}  & 0 & 0 \\
                 0 & B_{ab} & 0 \\
                 0 & 0 & B_{c}  \\
               \end{array}
             \right),
\end{equation}
and
\begin{eqnarray}
K_{ab} &=& \left(A_{ab}\langle\hat{S}_c^{a}\rangle +
B_{ab}\langle\hat{S}_f^{ab}\rangle\right)/H_0\\
K_c &=& \left(A_{c}\langle\hat{S}_c^{c}\rangle +
B_c\langle\hat{S}_f^{c}\rangle\right)/H_0,
\end{eqnarray}
so for $T>T^*$,
\begin{eqnarray}
\label{eqn:KSM}
K_{ab} &\approx& 2B_{ab}\chi_{ab}(T)\\
\nonumber K_c &\approx& 2B_c\chi_{c}(T).
\end{eqnarray}
In the interest of notational simplicity, we have not included any
indices on the on-site and transferred hyperfine coupling constants,
$A_{\alpha}$ and $B_{\alpha}$ for the different sites (In(1), In(2)
and M).  In general, each one of these couplings is different, and
is material dependent.

\section{The Knight Shift Anomaly}

In a generic Kondo lattice system, there are two spin species,
$\hat{S}_c$ and $\hat{S}_f$.  Therefore, there are three distinct
spin susceptibilities: $\chi_{cc}=\langle
\hat{S}_c\hat{S}_c\rangle$, $\chi_{cf}=\langle
\hat{S}_c\hat{S}_f\rangle$, and $\chi_{f\!f}=\langle
\hat{S}_f\hat{S}_f\rangle$.  In Eqs. \ref{eqn:KSIn1},
\ref{eqn:KSIn2} and \ref{eqn:KSM} we made the approximation that
$\langle S_c \rangle \ll \langle S_f \rangle$ for $T\gg T^*$.  A
more complete description is given by $g_c\mu_B\langle S_c \rangle =
(\chi_{cc}(T) + \chi_{cf}(T)) H_0$ and $g_f\mu_B\langle S_f \rangle
= (\chi_{ff}(T) + \chi_{cf}(T)) H_0$, where $g_{c,f}$ are the
g-factors.  In fact, the full expression for the Knight shift is
given by:
\begin{eqnarray}
K_{\alpha}^{(\eta)}(T) &=&
\left(A_{\alpha}^{(\eta)}\right)\chi_{cc,\alpha}(T) +
\left(A_{\alpha}^{(\eta)} + \sum_{i\in
nn}B_{\alpha}^{(\eta)}\right)\chi_{cf,\alpha}(T) +\\
\nonumber&&\left(\sum_{i\in nn}
B_{\alpha}^{(\eta)}\right)\chi_{ff,\alpha}(T),
\end{eqnarray}
where $\eta$ refers the the particular nuclear site, $\alpha$ is the
field direction, and  for simplicity we have absorbed the g-factors
into the definition of the hyperfine constants
\cite{twocomponentNMR}. The bulk susceptibility is given by:
\begin{equation}
\chi_{\alpha}(T) = \chi_{cc,\alpha}(T) + 2\chi_{cf,\alpha}(T) +
\chi_{ff,\alpha}(T).
\end{equation}
Note that if $A_{\alpha}^{(\eta)}=\sum_{i\in nn}
B_{\alpha}^{(\eta)}$, then the relation $K\propto\chi$ is recovered,
however in general this relation will not hold.  If $\chi_{cc}(T)$,
$\chi_{cc}(T)$, and $\chi_{cc}(T)$ have different temperature
dependences, then the Knight shift will not be proportional to
susceptibility, leading to a Knight shift anomaly at a temperature
$T^*$.

Since a complete solution of the Kondo lattice remains a challenge
for theory at present, there is no description of the temperature
dependence of $\chi_{cc}(T)$, $\chi_{cc}(T)$, and $\chi_{cc}(T)$.
However, using empirical observations Nakatsuji and coworkers
developed  a phenomenological two-fluid picture of the
susceptibility and specific heat in \cecoin\ which also works well
for understanding the Knight shift anomaly
\cite{NFP,twocomponentNMR,curroMRS}. In this picture, $\chi_{cf}(T)
\sim \left(1-T/T^*\right)\log(T^*/T)$ for $T<T*$, and $\chi_{cf}(T)
\sim 0$ for $T>T^*$, and we assume $\chi_{cc}$ is negligible for all
temperatures. This relation appears to hold for several different
heavy fermion and mixed valent compounds down to the relevant
ordering temperatures \cite{twocomponentNMR,curroSCES}. Recently we
have found evidence that $\chi_{cf}(T)$ saturates below a
temperature $T_0 \ll T^*$ in \cecoin\ for $H>H_{c2}$.  Recent
theoretical work lends support for the two-fluid picture
\cite{pingsun,barzykin}, but a full description is still lacking.
Nevertheless, the NMR data in a wide range of materials strongly
suggest that a complete theoretical description should involve one
localized degree of freedom giving rise to $\hat{S}_f$, and another
delocalized degree of freedom, giving rise to $\hat{S}_c$.

\section{The Spin Lattice Relaxation Rate and the Form Factors}

The spin lattice relaxation of the nuclei in the \cemin\ materials
is dominated by magnetic fluctuations mediated by the dynamic
hyperfine fields at the nuclear sites.  The \slrrtext, \slrr, can be
written in general as:
\begin{equation}
\label{eqn:t1general} \frac{1}{T_{1,{\alpha}}}
=\gamma_n^2\sum_{\beta}\int_0^{\infty}{\langle
H_{\beta}(t)H_{\beta}(0)\rangle e^{i\omega_L t}}dt,
\end{equation}
where we have dropped the "hyp" notation for simplicity, $\omega_L$
is the Larmor frequency, $\gamma_n$ is the nuclear gyromagnetic
ratio, and the sum is over the two spatial directions orthogonal to
the $\alpha$ direction. In general, the hyperfine field is given by:
\begin{equation}
\mathbf{H}(t)=\mathbb{A}\cdot\hat{\mathbf{S}}^c(0,t)+\sum_{i\in\mathrm{
nn}}\mathbb{B}^{(i)}\cdot\hat{\mathbf{S}}^f(\mathbf{r}_i,t)
\end{equation}
where  the $i$-sum is over the nearest neighbor Ce spins, and the
$\gamma$ sum is over the spatial directions.  This expression can be
simplified by writing:
\begin{equation}
H_{\beta}(t)=\sum_{\mathbf{q},\gamma}F^c_{\beta\gamma}(\mathbf{q})S^c_{\gamma}(\mathbf{q},t)+F^f_{\beta\gamma}(\mathbf{q})S^f_{\gamma}(\mathbf{q},t),
\end{equation}
where
$S^{(c,f)}_{\beta}(\mathbf{q},t)=\sum_{\mathbf{r}}\hat{S}_{\beta}^{(c,f)}(\mathbf{r},t)e^{-i\mathbf{q}\cdot\mathbf{r}}$,
and the form factors $\mathbb{F}^{(c,f)}(\mathbf{q})$ are given by:
\begin{eqnarray}
F^c_{\beta\gamma}(\mathbf{q}) & = & A_{\beta\beta}\delta_{\beta\gamma}\\
F^f_{\beta\gamma}(\mathbf{q}) & = & \sum_{i}
B_{\beta\gamma}^{(i)}e^{i\mathbf{q}\cdot\mathbf{r}_i}.
\end{eqnarray}
We then have:
\begin{eqnarray}
\nonumber\langle H_{\beta}(t)H_{\beta}(0) \rangle &=&
\sum_{\mathbf{q},\gamma}
\left(F_{\beta\gamma}^c(\mathbf{q})\right)^2 \langle
S_{\gamma}^{(c)}(\mathbf{q},t)S_{\gamma}^{(c)}(\mathbf{q},0)
\rangle+\\
\nonumber&&
2\sum_{\mathbf{q},\gamma}F_{\beta\gamma}^c(\mathbf{q})F_{\beta\gamma}^f(\mathbf{q})
\langle S_{\gamma}^{(c)}(\mathbf{q},t)S_{\gamma}^{(f)}(\mathbf{q},0)
\rangle+\\
&&\sum_{\mathbf{q},\gamma}
\left(F_{\beta\gamma}^f(\mathbf{q})\right)^2 \langle
S_{\gamma}^{(f)}(\mathbf{q},t)S_{\gamma}^{(f)}(\mathbf{q},0)
\rangle.
\end{eqnarray}
Here we have assumed that spin correlations at different
$\mathbf{q}-$vectors and directions are uncorrelated: $\langle
S_{\beta}^{(c,f)}(\mathbf{q},t)S_{\gamma}^{(c,f)}(\mathbf{q}',0)\rangle=\delta_{\mathbf{q}\mathbf{q}'}\delta_{\beta\gamma}
\langle
S_{\beta}^{(c,f)}(\mathbf{q},t)S_{\beta}^{(c,f)}(\mathbf{q},0)\rangle$,
although such an assumption may not be always valid \cite{Meier}.
Using the fluctuation-dissipation theorem, Eq. \ref{eqn:t1general}
can be rewritten as:
\begin{equation}
\label{eqn:t1twocomponent} \frac{1}{T_{1,\alpha}} =
\frac{1}{T_{1,\alpha}^{cc}}+\frac{2}{T_{1,\alpha}^{cf}}+\frac{1}{T_{1,\alpha}^{f\!f}},
\end{equation}
where:
\begin{equation}
\frac{1}{T_{1,{\alpha}}^{(\eta\eta')}} ={\gamma_n^2 k_B
T}\lim_{\omega\rightarrow
0}\sum_{\mathbf{q},\beta}\phi_{\beta}^{\eta\eta'}(\mathbf{q})\frac{\chi_{\eta\eta'\beta}''(\mathbf{q},\gamma)}{\hbar\omega},
\end{equation}
where $\chi_{\eta\eta'\beta}''(\mathbf{q},\omega)$ is the dynamical
spin susceptibility along the $\beta$ direction, and
\begin{equation}
\label{eqn:filterfunction}
\phi_{\beta}^{\eta\eta'}(\mathbf{q})=\sum_{\gamma}F_{\beta\gamma}^{(\eta)}(\mathbf{q})F_{\beta\gamma}^{(\eta')}(\mathbf{q}).
\end{equation}

Clearly, the behavior of the spin lattice relaxation predicted by
Eq. \ref{eqn:t1twocomponent} can be quite complex, depending on the
temperature dependence of the various dynamical spin
susceptibilities $\chi_{cc}''$, $\chi_{cf}''$ and $\chi_{f\!f}''$,
especially if the temperature dependence of these three quantities
differ. At present, there is no clear theoretical model for how
these susceptibilities should behave.   However, we can make several
important and general conclusions. NMR Knight shift studies and
theoretical considerations suggest that $\chi_{cf}$ is negligible
for $T>T^*$ \cite{CurroAnomalousShift}. Therefore, we expect that
$T_1^{-1}\approx T_{1,cc}^{-1} + T_{1,f\!f}^{-1}$ for $T> T^*$.
Below $T^*$, a new contribution, $T_{1,cf}^{-1}$ can become
important, and may lead to an anomalous temperature dependence, as
observed in \cecoin\ for fields applied in the $ab$ plane
\cite{curroPRL}.

The filter functions $\phi_{\beta}^{\eta\eta'}(\mathbf{q})$ given in
Eq. \ref{eqn:filterfunction} enhance or suppress the relative
importance of spin fluctuations to relax the nuclei at different
parts of the Brillouin zone.   For the In(1) we have:
\begin{eqnarray}
\nonumber\phi_{a,b}^{f\!f}(\mathbf{q})&=&16B_0^2\cos^2(q_xa/2)\cos^2(q_ya/2)\\
&&+16B_a^2\sin^2(q_xa/2)\sin^2(q_ya/2)\\
\phi_{c}^{f\!f}(\mathbf{q})&=&16B_z^2\cos^2(q_xa/2)\cos^2(q_ya/2)
\end{eqnarray}
Note that $F_{\beta}^{c}(\mathbf{q}) = A_{\beta\beta}$ is
$\mathbf{q}-$independent. For the In(2),
\begin{eqnarray}
\nonumber\phi_{a}^{f\!f}(\mathbf{q})&=&4(B_{0} + B_a\cos{2\theta})^2\cos^2(q_xa/2)+\\
&&4(B_a\sin{2\theta})^2\sin^2(q_xa/2)\\
\phi_{b}^{f\!f}(\mathbf{q})&=&4B_y^2\cos^2(q_xa/2)\\
\nonumber\phi_{c}^{f\!f}(\mathbf{q})&=&4(B_{0} - B_a\cos{2\theta})^2\cos^2(q_xa/2)+\\
&&4(B_a\sin{2\theta})^2\sin^2(q_xa/2)
\end{eqnarray}
and for the M site:
\begin{eqnarray}
\phi_{a}^{f\!f}(\mathbf{q})&=&4B_{ab}^2\cos^2(q_zc/2)\\
\phi_{c}^{f\!f}(\mathbf{q})&=&4B_c^2\cos^2(q_zc/2)
\end{eqnarray}
These quantities are plotted in Figs. \ref{fig:phiplot1} and
\ref{fig:phiplot}. The anisotropic nature of the hyperfine tensor of
the In(1) and In(2) gives rise to a second component in their filter
functions that does not vanish at $\mathbf{q} = (\pi/a,\pi/a,q_z)$.
A priori, one might expect that the In(1) and In(2) would be
insensitive to fluctuations at these wavevectors because they are
located in symmetric positions in the unit cell with respect to the
Ce. However, since the bond axes are not coincident with the unit
cell axes, the hyperfine tensor is not diagonal.  As a result, the
In(1) and In(2) are sensitive to antiferromagnetic fluctuations
\cite{curroPRL}.  In the cuprates, there is no such anisotropy in
the hyperfine tensors, since the bond axes of the Cu and the O
nuclei lie along the unit cell directions.

\begin{figure}
\begin{centering}
\includegraphics[width=0.7\linewidth]{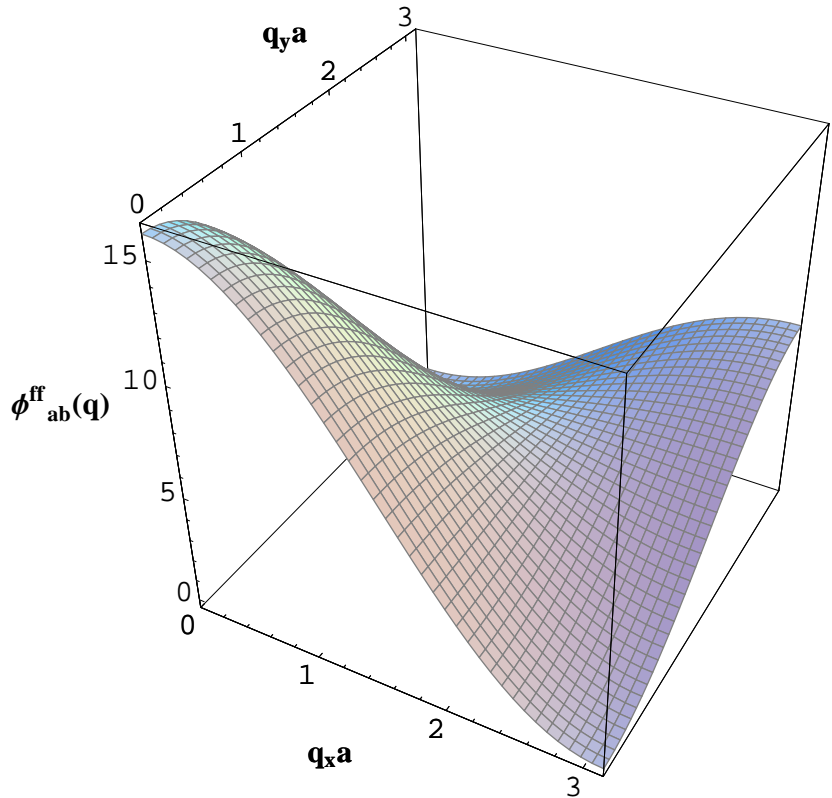}
\caption{\label{fig:phiplot1} The filter function
$\phi_{ab}^{f\!f}(\mathbf{q})$ for the In(1), shown for $B_a/B_0 =
0.6$. Note that $\phi_{ab}^{f\!f}(\mathbf{q})$ remains finite for
$\mathbf{q} = (\pi/a,\pi/a,q_z)$.}
\end{centering}
\end{figure}

\begin{figure}
\begin{centering}
\includegraphics[width=\linewidth]{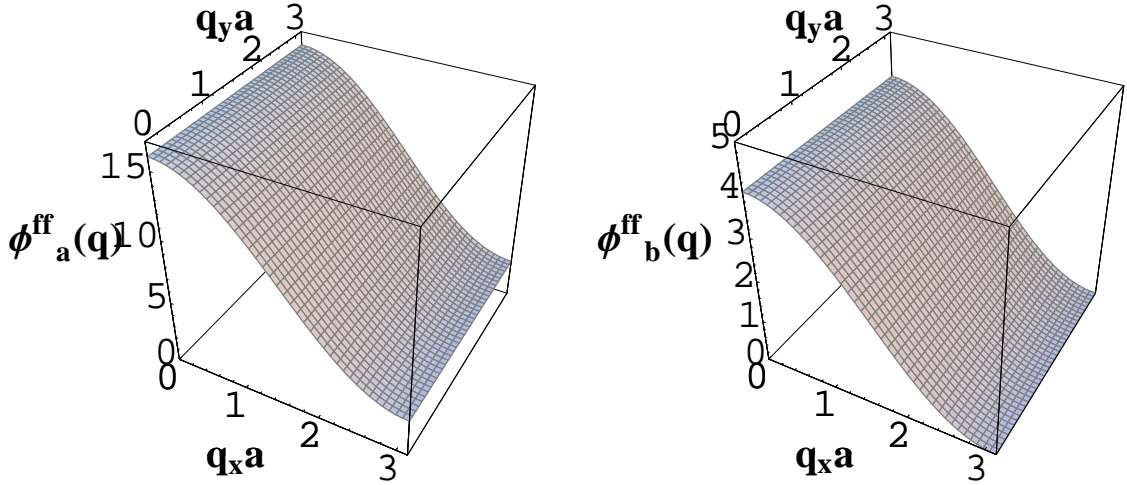}
\caption{\label{fig:phiplot} The filter functions
$\phi_{a}^{f\!f}(\mathbf{q})$ and $\phi_{b}^{f\!f}(\mathbf{q})$ for
the In(2), shown for $B_a/B_0 = 0.3$. Note that
$\phi_{a}^{f\!f}(\mathbf{q})$ remains finite for all $\mathbf{q}$,
whereas $\phi_{b}^{f\!f}(\mathbf{q})$ does not.}
\end{centering}
\end{figure}

It should be pointed out that the the spin lattice relaxation at the
In(2) site is particularly difficult to interpret, because the
electric field gradient (EFG) does not have axial symmetry.  The
quadrupolar part of the hamiltonian of the In(2) is given by:
\begin{equation}
\hat{\mathcal{H}}_Q = \frac{h\nu_{cc}}{6}\left[
(3\hat{I}_c^2-\hat{I}^2) +
\frac{\nu_{aa}-\nu_{bb}}{\nu_{cc}}(\hat{I}_a^2-\hat{I}_b^2)\right]
\end{equation}
where the $\nu_{\alpha\alpha}$  are the eigenvalues of the EFG
tensor \cite{CPSbook}. Since $\hat{\mathcal{H}}_Q$ does not commute
with $\hat{I}_a$, $\hat{I}_b$ or $\hat{I}_c$, the eigenstates of the
nuclear levels are non-trivial superpositions of the $|m\rangle$
states, and the usual selection rules for transitions between
nuclear levels do not hold. Consequently, unlike the In(1) which
does have axial symmetry, the master equation which governs the
relaxation of the In(2) nuclei cannot be solved analytically.  One
can measure the relaxation at the In(2) site, which is a function of
the rates given in Eq. \ref{eqn:t1general}, but it is not
straightforward to independently measure the rates
$T_{1,\alpha}^-1$.

In fact, several groups have measured the temperature dependence of
the relaxation rate at the In(1) site in the various 115 compounds,
and have found a rich variety of behavior both in the
superconducting and normal states \cite{zhengpressure,zhengQCP,
koharaambient,koharapressure,zhengalloys}.  In \cerhin, the
temperature dependence of the  spin lattice relaxation is a strong
function of pressure \cite{zhengpressure}, and even shows signs
reminiscent of the pseudogap observed in the high temperature
superconductors.  One explanation is that the spin fluctuations are
modified with pressure as the ground state evolves from
antiferromagnetic to superconducting.  However, the internal field
in the ordered state is reduced linearly with pressure, while the
ordered moment remains constant \cite{anna}.  This result suggests
that the hyperfine constants may change with pressure, a result that
has not yet been confirmed by Knight shift measurements under
pressure. If the hyperfine constants are pressure dependent, then
the temperature dependence of the \slrr\ dominantly reflects these
changes, rather than changes in the spin fluctuation spectrum.
Measurements of \slrr\ in CeRh$_{1-x}$Ir$_x$In$_5$ alloys and in
CeRhIn$_5$ under pressure suggest that there are regions of the
phase diagram where both antiferromagnetism and superconductivity
coexist microscopically \cite{zhengpressure,zhengalloys}.  In such
cases, it appears that the typical $T^3$ temperature dependence in
the superconducting state of d-wave superconductors may be modified,
possibly due to the presence of the antiferromagnetic order. Further
studies of the Knight shift in these systems may shed important new
light on this behavior.

\section{NMR in the Actinide 115's}

Recently the actinide based 115's, synthesized with U, Np and Pu
have attracted attention due to the large superconducting transition
temperature in \pucoga\ \cite{pucoga5discovery,curropucoga5nature}.
One of the striking discoveries in \pucoga\ is that the temperature
dependence of \slrrtext\ scales with the transition temperature in
the same manner in \cecoin, \pucoga, and YBa$_2$Cu$_3$O$_7$
suggesting a common mechanism. Although there is little detailed
information available about the hyperfine couplings present in the
\pumga\ materials, there is no reason to suspect that the general
form of the tensors $\mathbb{A}$ and $\mathbb{B}$ should differ from
those in the \cemin\ materials.  Therefore, the \slrrtext\
measurements, whether on the In(1), In(2), Ga(1) or Ga(2), should be
sensitive to antiferromagnetic fluctuations at $(\pi/a,\pi/a,q_z)$.
The unusual temperature dependence of \slrr\ may well reflect the
temperature dependence of $\chi_{f\!f}''$ and $\chi_{cf}''$.  The
\slrrtext\ measured in PuRhGa$_5$, however, differs dramatically
from that found in \pucoga\ below $\sim$30K, well above \tc\
\cite{sakai}. Although this anomaly may reflect unusual spin
dynamics \cite{yunkyu}, another explanation may be the presence of a
Knight shift anomaly at 30K observed in this material
\cite{walstedtprivate}.  This discrepancy highlights the unusual
hyperfine interactions present in all of the heavy fermion
materials, and demonstrates the need for a general understanding of
the two spin components and their dynamics before placing too much
emphasis on the behavior of the spin lattice relaxation as a
function of temperature.

\section{Acknowledgments}

We thank D. Pines, J. Schmalian, and C. P. Slichter for enlightening
discussions. This work was performed at Los Alamos National
Laboratory under the auspices of the U.S. Department of Energy.

\end{document}